\documentclass[%
reprint,
superscriptaddress,
twocolumn,
nofootinbib,
notitlepage,
 amsmath,amssymb,
aip,
showkeys,
apl,
]{revtex4-1}

\usepackage{graphicx}

\usepackage{dcolumn}
\usepackage{bm}
\usepackage{hyperref}
\usepackage{color}

\usepackage[utf8]{inputenc}
\usepackage{upgreek}
\usepackage{relsize}
\usepackage{footmisc}

\allowdisplaybreaks[1]

\begin{document}


\title{High charge plasma electron bunches from two collinear laser pulses }

\author{Vojtěch Horný}
\email{horny@ipp.cas.cz}
\affiliation{Institute of Plasma Physics, Czech Academy of Sciences, Za Slovankou 1782/3, 182 00 Praha 8, Czech Republic}
\affiliation{Faculty of Nuclear Sciences and Physical Engineering, Czech Technical University in Prague, Břehová 7, 115 19 Praha 1, Czech Republic}
\author{Ondřej~Klimo}
\affiliation{Faculty of Nuclear Sciences and Physical Engineering, Czech Technical University in Prague, Břehová 7, 115 19 Praha 1, Czech Republic}
\affiliation{Institute of Physics, Czech Academy of Sciences, Na Slovance 1999/2, 182 21, Praha 8, Czech Republic}
\author{Miroslav~Krůs}
\affiliation{Institute of Plasma Physics, Czech Academy of Sciences, Za Slovankou 1782/3, 182 00 Praha 8, Czech Republic}
\affiliation{Institute of Physics, Czech Academy of Sciences, Na Slovance 1999/2, 182 21, Praha 8, Czech Republic}

\date{\today}

\begin{abstract}
An optical injection scheme into the laser wakefield accelerator by preceding injection pulse is investigated by means of 3D numerical particle-in-cell simulations. Quasimonoenergetic hundred-pC electron bunches as short as 6~fs can be generated. Optimal beam separation distance  is found at the intersection point of the injection beam bubble with the collection volume for transverse injection into the accelerator beam bubble. It approximately corresponds to the plasma wavelength. The main advantage of this scheme is the localized injection of high charge. This injection mechanism can be useful for applications such as ultrashort and relatively intense X-ray radiation sources such as a betatron radiation or Thomson backscattering, time-resolved electron diffraction or for seeding of further acceleration stages.
\end{abstract}


\maketitle

The non-linear regime of laser wakefield acceleration (LWFA) is a promising method to generate high energy electron bunches by an interaction of an ultra-short (a few tens of fs) ultra-intense ($I>10^{19}$~W/cm$^2$) laser pulse with a gaseous target. The main advantage of this plasma-based concept is in the ability of plasma to sustain large accelerating gradients of the order of hundreds of GV/m\citep{esarey2009physics}. This is almost three orders of magnitude higher than the field in conventional radiofrequency accelerators. Therefore, GeV electron bunches can be generated within cm acceleration distance.  Nevertheless, there are still major issues which have to be solved in order to make such method practically feasible, e.g. as a driver of compact X-ray free electron lasers. 

The most simple mechanism to inject plasma electron into the accelerating stage of the non-linear wakefield is self-injection\citep{kostyukov2004phenomenological}. However, this process is of unstable, nonlinear nature. As a consequence, it is very difficult to control parameters of produced electron beams. Several controlled injection schemes were suggested, such as density down ramp injection \citep{geddes2008plasma, schmid2010density}, ionisation injection \citep{pak2010injection, clayton2010self, mirzaie2015demonstration}, and various optical injection mechanisms with perpendicularly crossed \citep{umstadter1996laser, wang2008controlled, horny2017short} or counter-propagating\cite{esarey1997electron, fubiani2004beat, kotaki2004head, davoine2009cold, lehe2013optical} colliding pulses. 

Alternative optical injection configurations use two co-propagating pulses in two different configurations. In the first approach, a more intense laser pulse is delayed to provide injection by an optically induced ionization in the linear or moderately non-linear LWFA regimes \citep{bourgeois2013two, xu2014low}.  Both simulation works report few pC bunches with low energy spread and emittance due to localized ionization injection by tightly focused injection pulse. 

The second approach does not rely on optical field ionization, but it is based on different focusing of both collinear pulses. Thomas \textit{et al.} \cite{thomas2008monoenergetic}  used laser pulse with a large focal spot to generate plasma density perturbation as a guiding structure for a tightly focused pulse. They even measured quasimonoenergetic electron beams. 

Hu \textit{et al.} \citep{hu2016brilliant} recently proposed injection by tightly focused intense pulse following the plasma wave drive pulse at sharp vacuum-plasma transition shorter than ten microns. It is challenging to be realized such sharp density transitions in experiments using state-of-art technology though.  Few pC GeV electron beams with energy spread under 0.5~\% and sub-$\upmu$m normalized transverse emittance are predicted employing the phase-space rotation with the rephasing technique \cite{yang2006femtosecond}.

The aforementioned injection techniques are usually aimed on the optimizing of a single specific electron bunch parameter such as its charge, energy spread or emittance. However, this Letter proposes a scheme, where high charge electron bunches with relatively low energy spread and emittance are generated. It is achieved by an optical injection scheme with a weaker injection pulse preceding the accelerating plasma wave driving pulse. 
 Let us note that similar idea is also independently suggested by another group \cite{zahra}.


This simple configuration avoids the issues with temporal and spatial synchronization which are characteristic for other optical injection schemes.  Contrary to previous schemes with collinear pulses \citep{bourgeois2013two, xu2014low} where the injection is longitudinal, the presented injection process is transverse and it leads to much higher charge.

 \begin{figure}[htbp]
   \includegraphics[scale=0.55]{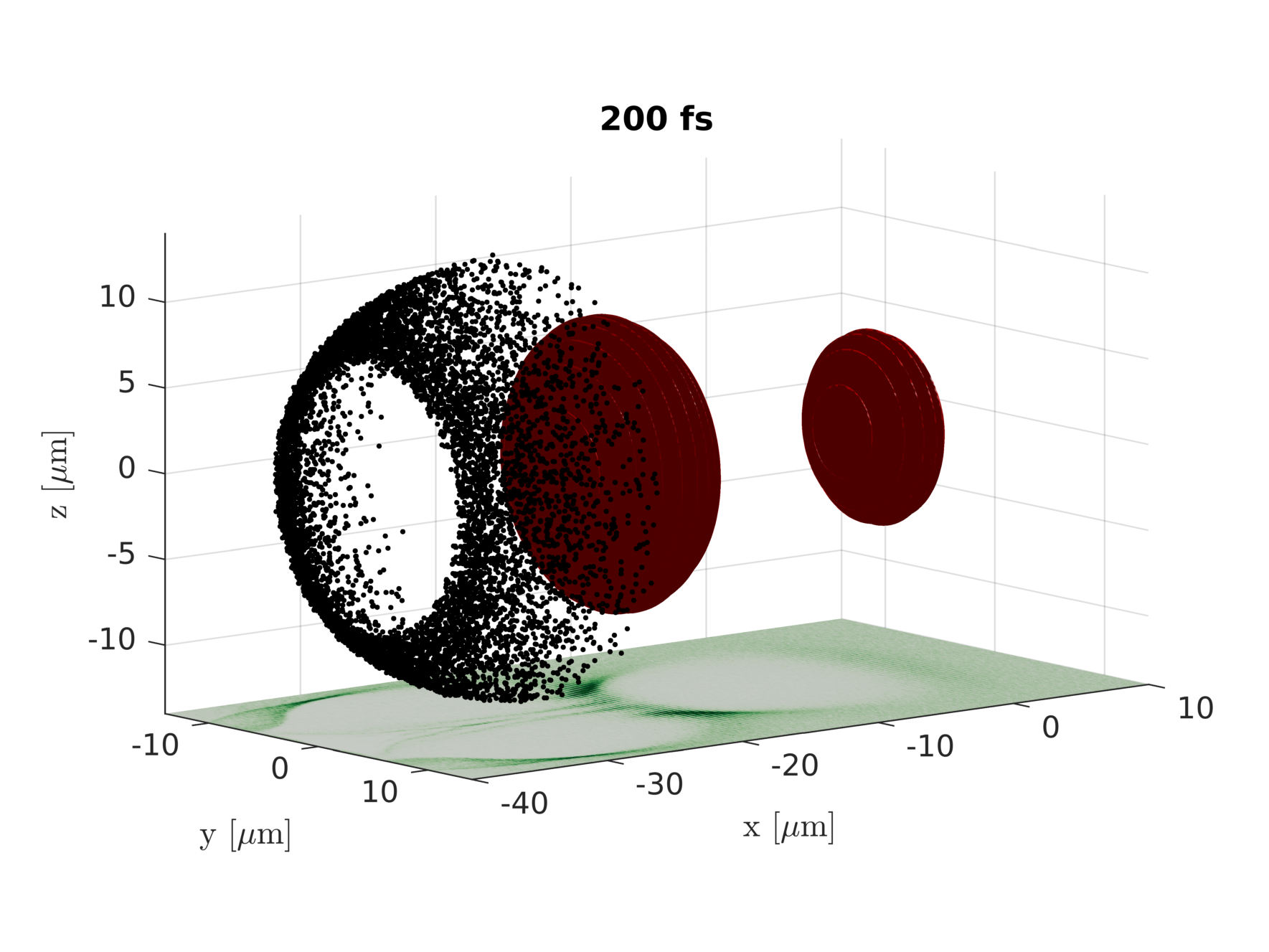}\\
      \includegraphics[scale=0.55]{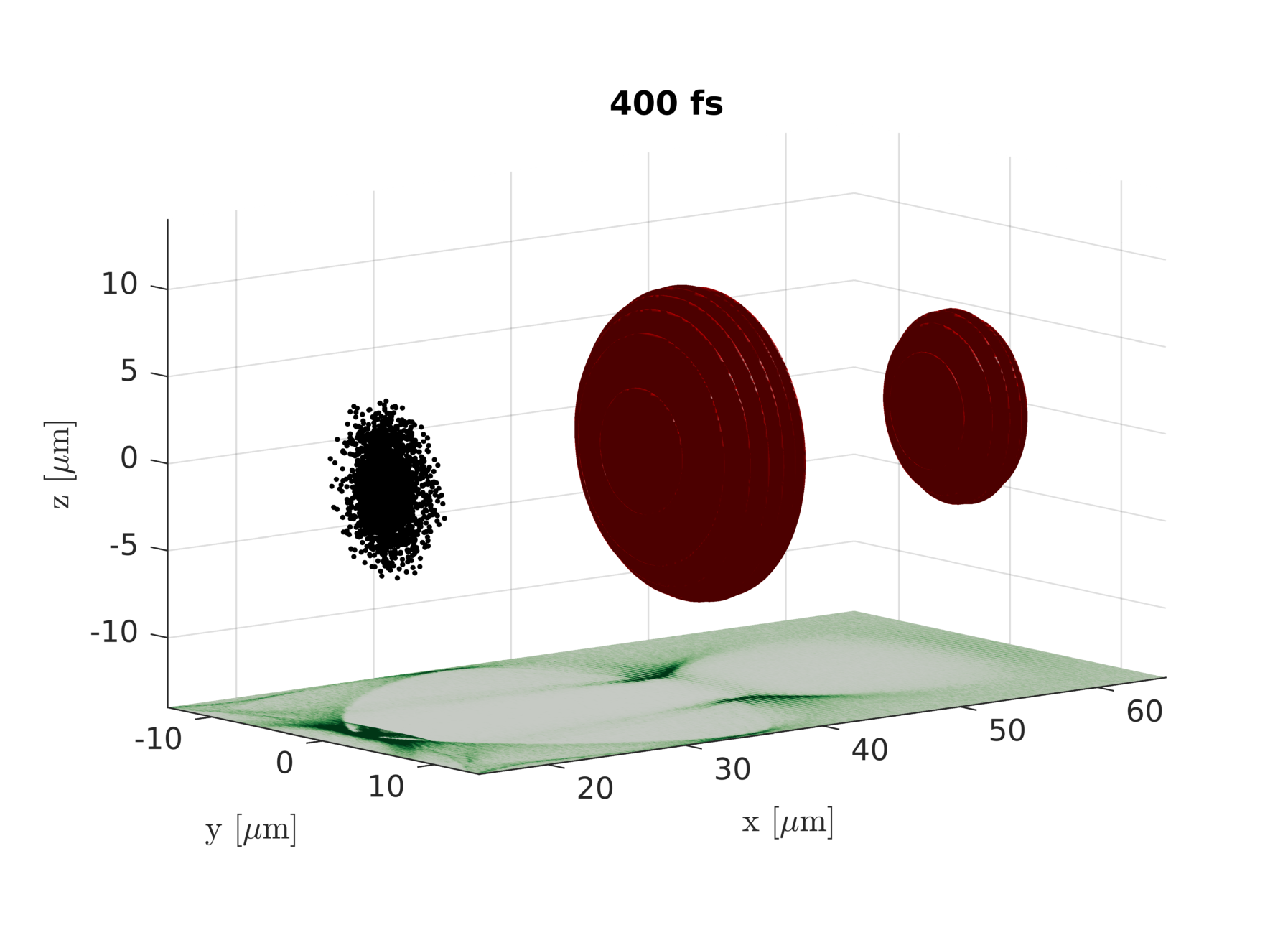}
   \caption{Visualization of the injection process from 3D PIC simulation. Isosurfaces of laser pulse electric field are displayed in brown. Trapped electrons (black dots, selection) are initially at the ring around axis. The electron density at the plane $z=0$ is shown at the bottom of both boxes.}
\label{fig:evo}
\end{figure}


The collection volume for the transversely self-injected electrons in nonlinear bubble regime is the ring around laser propagation axis \citep{benedetti2013numerical}  with radius 
\begin{equation}
r_0 = k_p^{-1}(-2.0+1.4a_0-0.05a_0^2),
\end{equation}
where $k_p=\omega_p/c$, $c$ is speed of light in vacuum, $\omega_p = \sqrt{n_ee^2/m_e\varepsilon_0}$ is plasma electron frequency, $n_e$ is electron density, $e$ is electron charge, $m_e$ is electron mass, $\varepsilon_0$ is vacuum permittivity, and $a_0\simeq 0.855\times 10^{-9} \sqrt{I_0 \textnormal{ [W/cm$^2$]}}\lambda \textnormal{ [$\upmu$m]}$ is the peak normalized vector potential of the laser, $I_0$ is peak intensity of the laser and $\lambda$ is laser wavelength. The presented injection scheme provides higher injected charge by increasing the number of plasma electrons in such a ring. This enables generation of 180~pC bunches with energy spread of 9\% and normalized transverse emittance of few $\upmu$m for feasible experimental conditions.

In this injection scheme, the intensity of the injection pulse is assumed to be high enough ($a_{1}\gtrsim 1.8$) to generate its own bubble. The longitudinal radii of drive and injection bubbles are\citep{benedetti2013numerical}
\begin{equation}
R_{\parallel,0,1} = k_p^{-1}(2.9+0.305a_{0,1}).
\end{equation}
Drive pulse delay is chosen in such a way that the collection ring for self-injection just in front of it coincides with the electron sheath of the first bubble, i.e.
\begin{equation}
\Delta t = \frac{1}{c}(\sqrt{R_{\parallel,0}^2-r_0^2}+\sqrt{R_{\parallel,1}^2-r_0^2}) + \frac{\tau}{2},
\label{eq:delay}
\end{equation}
where $\tau$ pulse duration (FWHM of intensity). Such configuration increases the electron density in a region where electrons could be potentially injected to bubble dragged by main pulse. 

Nevertheless, it is still not a sufficient condition to induce an injection, which can be also understood as a Langmuir wave-breaking. Lehe et al. \citep{lehe2013optical} showed that the wave-breaking can be induced by the triggered expansion of the bubble as a whole. Within our scheme, bubble expansion can also occur due to stochastic nature of the bubble dynamics, but such scheme would not be stable.

Therefore, the wave breaking is achieved in a controlled manner at a density up-ramp at a vacuum plasma transition, similarly as by the up-ramp injection by a single pulse in much higher plasma densities \cite{li2013dense}. Such localized injection leads to quasimononenergetic electron spectra and potentially to a good reproducibility.

The injection pulse also modulates the electron density in the location where the main pulse propagates. Thus, in some sense,  this letter follows on previous research on plasma waveguide \cite{borisov1994stable, chen1998evolution}.

 The injection process was studied by means of 3D particle-in-cell (PIC) simulations using the code EPOCH \cite{arber2015contemporary}. The following parameters were chosen to demonstrate the scheme: laser wavelength $\lambda=0.8$~$\upmu$m, waist size $w_0=9.5$~$\upmu$m, pulse duration $\tau=25$~fs, drive and injection laser pulses strength parameters  $a_{0}=4$ and  $a_{1}=2.5$. The mutual delay between pulses was 65~fs which corresponds to plasma wavelength, both pulses are linearly polarized. A homogeneous electron gas with density $3\times 10^{18}$~cm$^{-3}$ and immobile ions were assumed (they were not simulated). The initial demonstration used 20~$\upmu$m long linear front plasma density ramp. Simulation box dimensions were 85~$\upmu$m$\times$36~$\upmu$m$\times$36~$\upmu$m with 25$\times$4$\times$4 cells per wavelength and 2 particles per cell. 
  
 The snapshots of the injection process from the 3D PIC simulation are shown in Figure \ref{fig:evo}. The injected electrons lying initially at the ring around the propagation axis located at the transition between the end of the density ramp and the homogeneous plasma are at first disturbed by the injection pulse and after that trapped in the bubble dragged by the main pulse. The nature of this injection process is transverse. The incline of the density ramp tunes the injected charge; the shorter density ramp leads to the higher charges. This behaviour was studied for feasible ramp lengths of 20--100~$\upmu$m.
  
  \begin{figure}
    \includegraphics[scale=1.0]{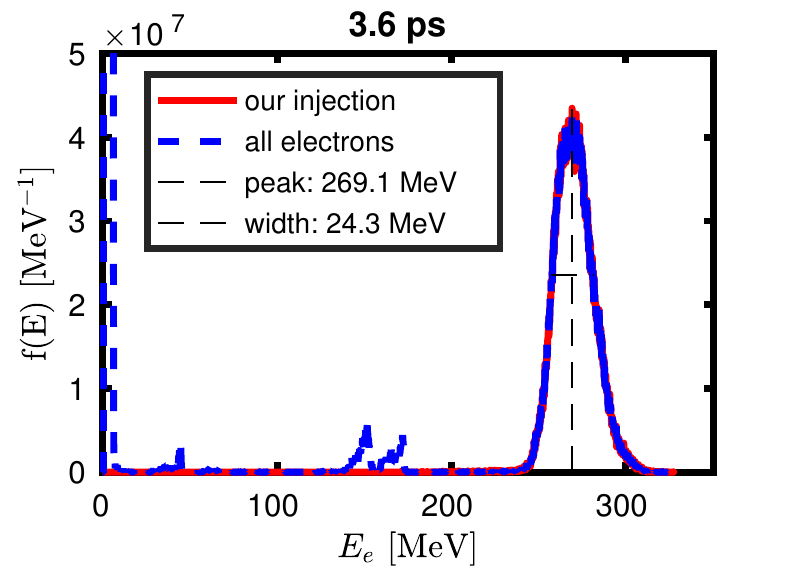}
    \caption{Electron spectrum after 3.6~ps of acceleration from 3D simulation.
    }
    \label{fig:enspec}
\end{figure}

\begin{figure}
\includegraphics[scale = 1.0]{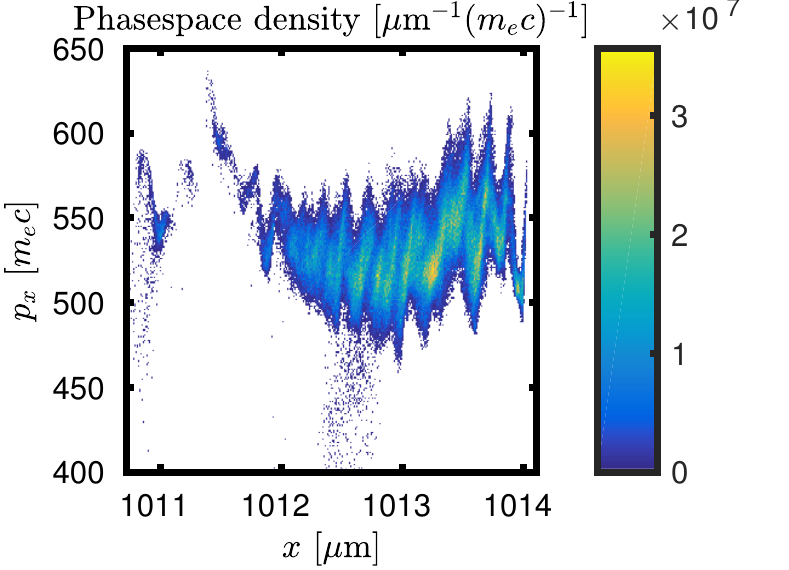}
\caption{Phasespace density of the accelerated electron bunch at time 3.6 ps.}
\label{fig:phasespace}
\end{figure}

Energy, relative energy spread, charge, and transverse emittance of the trapped electron bunch in the simulation were 269~MeV, 9~\%, 188~pC, and 1.63~$\pi\cdot$mm$\cdot$mrad, respectively, after 3.6~ps. The main feature of this scheme is that an electron bunch with a relatively large charge is acclerated to energies at least higher than 250~MeV, while simultanously having relatively low FWHM energy spread and acceptable value of emittance. Energy spectrum is depicted in Figure \ref{fig:enspec}. Let us stress that almost no dark current is generated. It means that the bunch quality achievable with this injection scheme may be in certain aspects better than by other injection schemes, even though these may claim e.g. lower energy spread. The longitudinal  phase-space plot is shown in Figure \ref{fig:phasespace}. The bunch length is 1.8~$\upmu$m (FWHM of bunch density along $x$-axis). Energy spread could be reduced by intricate density tailoring techniques in a further stage of acceleration in the leading plasma channel up to multi-GeV energies, similarly as in \citep{hu2016brilliant}. Such electron bunch could be also used as an intense source of few-fs long hard X-ray pulses of betatron radiation as it was shown in \cite{horny2017temporal}.

Additional 3D simulations were performed in order to determine the dependence of the accelerated electron bunch parameters on the length of the density ramp and on the intensity of the injection pulse. The results were compared at the time of 2~ps of simulation when the injection process is already finished and acceleration stage is stabilized.  It was observed, that for given parameters, the ramp length of 20~$\upmu$m is the optimal one from the point of view of the highest injected charge $Q$, while energy spread $\Delta E$ (FWHM of energy peak) is kept as low as 9\%. Such ramp length is also feasible from the experimental point of view. The dependence is presented in Table \ref{tab:ram}. 

Table \ref{tab:int} illustrates that there is a wide range of injection pulse intensities which lead to high charge electron bunches. Nevertheless, the optimum parameters are achieved when injection pulse intensity is high enough to generate its own wake, i.e. $a_{1} \gtrsim 1.8$. However, too strong injection pulse can destroy wakefield driven by drive pulse, i.e.  and $(a_{1}/a_{0})^2 < 0.5$.  Our injection mechanism was not observed for $a_{1} \gtrsim 3.5$;  self-injection occurred in the first bubble, however, injected bunch was soon scattered by the main pulse located at the rear side of the first bubble.

\begin{table}
\caption{Dependence of electron bunch parameters on density ramp length  at time 2~ps of simulation. $l_r$ is initial linear ramp length, $E$ is energy of a peak of electron spectra, $\Delta E$ is its width, $Q$ is injected charge. }
\label{tab:ram}
\begin{tabular}{cccccc}
\hline
\hline
$a_{0}$ & $a_{1}$ & $\l_r$ [$\upmu$m]  & $E$ [MeV] & $\Delta E$ [MeV] & $Q$ [pC]  \\
\hline
4 & 2.5 & 10 & 163 & 12 & 182 \\
4 & 2.5 & 20 & 160 & 14 & 188 \\
4 & 2.5 & 30 & 162 & 20 & 167 \\
4 & 2.5 & 50 & 175 & 22 & 128 \\
4 & 2.5 & 100 & 154 & 50 & 75 \\
\hline
\hline
\end{tabular}
\end{table}
 
 \begin{table}
\caption{ Dependence of electron bunch parameters on injection pulse intensity at time 2~ps of simulation. $l_r$ is initial linear ramp length, $E$ is energy of a peak of electron spectra, $\Delta E$ is its width, $Q$ is injected charge.}
\label{tab:int}
\begin{tabular}{cccccc}
\hline
\hline
$a_{0}$ & $a_{1}$ & $\l_r$ [$\upmu$m]  & $E$ [MeV] & $\Delta E$ [MeV] & $Q$ [pC]  \\
\hline
4 & 1 & 30 & 227 & 25 & 39 \\
4 & 2 & 30 & 218 & 23 & 122 \\
4 & 2.5 & 30 & 162 & 20 & 167 \\
4 & 3 & 30 & 119 & 37 & 127 \\
\hline
\hline
\end{tabular}
\end{table}

Presented scheme is very sensitive to time delay between pulses. Its proper value was derived in \eqref{eq:delay}. It is 67~fs for our demonstration example. Table \ref{tab:del} shows that the parameters of accelerated electron bunches are optimal around this predicted value.

\begin{figure}
\centering
      \includegraphics[scale=1.0]{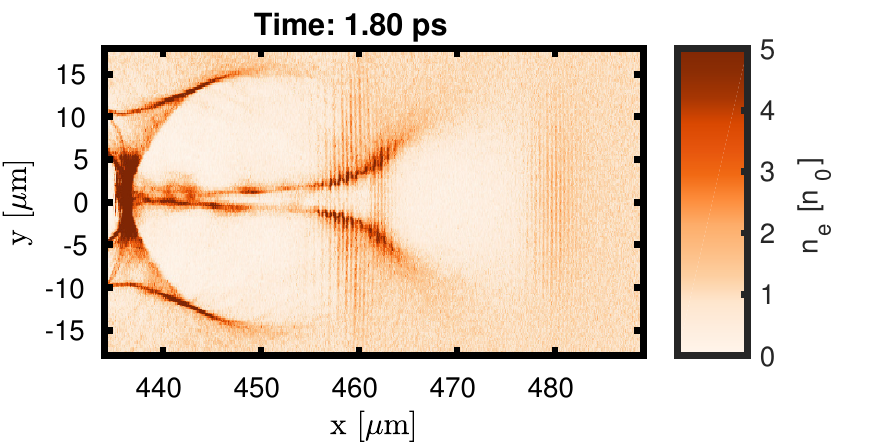}
   \caption{Disruption of the trapped electron bunch by the electron stream caused by the contact of the rear part of the first bubble and the drive pulse. Time delay between pulses was 75~fs.}
\label{fig:drawback}
\end{figure}

However, if the delay between pulses is too long, trapped electron bunch may be dispersed by an electron stream generated due to the contact between the most rear part of the injection pulse bubble and the drive pulse. Such phenomenon is displayed in Fig. \ref{fig:drawback} for the time delay of 75~fs. 

 \begin{table}
\caption{Dependence of electron bunch parameters on mutual delay between the both pulses $\tau$ at time 2~ps of simulation. $l_r$ is initial linear ramp length, $E$ is energy of a peak of electron spectra, $\Delta E$ is its width, $Q$ is injected charge. Values in italic at the time of 1 ps due to later collapse of acceleration, see Figure \ref{fig:drawback}.
When $\Delta \tau$ was set to 55~fs, injection occurred stochastically around the time of 1.8~ps. }
\label{tab:del}
\begin{tabular}{ccccccc}
\hline
\hline
$a_{0}$ & $a_{1}$ & $\Delta \tau$ [fs] & $\l_r$ [$\upmu$m]  & $E$ [MeV] & $\Delta E$ [MeV] & $Q$ [pC]  \\
\hline
4 & 2.5 &  60 &30 & 130 & 25 & 103 \\
4 & 2.5 &  65    & 30 & 162 & 20 & 167 \\
4 & 2.5 & 70   & 30 & \textit{97} & \textit{26} & \textit{140} \\
4 & 2.5 & 75 & 30 & \textit{97} & \textit{9} & \textit{81} \\
\hline
\hline
\end{tabular}
\end{table}


In conclusion, we suggested optical injection scheme of two mutually delayed laser pulses which provides the generation of quasimonoenergetic high charge electron bunches accelerated to hundreds-MeV energies by laser wakefield acceleration mechanism. Such scheme is applicable at 100-TW-class laser systems. Additionally, employing the second stage of acceleration in a plasma channel, multi-GeV bunches with sustained parameters can be generated. The main advantage of our scheme is that the generated electron bunch is simultaneously of high charge, short length, relatively low energy spread, and acceptable value of emittance. The parameters of preceding injection pulse and following the main pulse are the same but intensity. The idea of injection is based on the geometrical approach, and on the wave-breaking on up-ramp-plateau density transition.

\acknowledgments
Authors would like to thank Jiří Limpouch and Václav Petržílka for fruitful discussions and their valuable suggestions.
Authors wish to acknowledge financial support from Czech Science Foundation (GA ČR) project  15-03118S, Czech Technical University (CTU) project SGS16/248/\-OHK4/3T/14, and ELI - Extreme Light Infrastructure - phase 2 (CZ.02.1.01\-/0.0/0.0\-/15\verb!_!008/0000162) from European Regional Development Fund. The financial support provided by the Ministry of Education, Youth and Sports of the Czech Republic within the projects  LQ1606, and LD14089 is greatly appreciated. Access to computing and storage facilities owned by parties and projects contributing to the National Grid Infrastructure MetaCentrum provided under the programme \textit{Projects of Large Research, Development, and Innovations Infrastructures} (CESNET LM2015042), and to ECLIPSE cluster of ELI-Beamlines project is greatly appreciated as well.

\bibliography{IEEEabrv,2018_APL}

\end{document}